# Neptune's Dynamic Atmosphere from Kepler K2 Observations: Implications for Brown Dwarf Light Curve Analyses

Amy A. Simon[1], Jason F. Rowe[2], Patrick Gaulme[3], Heidi B. Hammel[4], Sarah L. Casewell[5], Jonathan J. Fortney[6], John E. Gizis[7], Jack J. Lissauer[8], Raul Morales-Juberias[9], Glenn S. Orton[10], Michael H. Wong[11], Mark S. Marley[8]

1. NASA Goddard Space Flight Center, Solar System Exploration Division (690.0), 8800 Greenbelt Road, Greenbelt, MD 20771, USA. Email: amy.simon@nasa.gov

2. Universite de Montreal, Departement de Physique, 2900 Boul. Edouard-Montpetit, Montreal, QC, Canada, H3T 1J4

3. New Mexico State University, Department of Astronomy, PO Box 30001, Las Cruces, NM 88003-4500, USA

4. AURA, Inc., 1212 New York Avenue NW, Washington, DC 20005, USA

5. University of Leicester, Department of Physics and Astronomy, University Road, Leicester, LE1 7RH, UK

6. University of California, Santa Cruz, Department of Astronomy & Astrophysics, 1156 High Street, 275 Interdisciplinary Sciences Building, Santa Cruz, CA 95064, USA

7. University of Delaware, Department of Physics and Astronomy, 104 The Green, Newark, DE 19716, USA

8. NASA Ames Research Center, Space Sciences & Astrobiology Division, MS 245-3, Moffett Field, CA 94035, USA

9. New Mexico Institute of Mining and Technology, Physics Department, Workman Center 345, 801 Leroy Place, Socorro, NM 87801, USA

10. Jet Propulsion Laboratory/California Institute of Technology, M/S 183-501, 4800 Oak Grove Drive, Pasadena, CA 91109, USA

11. University of California at Berkeley, Astronomy Department, Berkeley, CA 947200-3411, USA

Tables: 1
Supplemental online: 1 movie

Abstract

Observations of Neptune with the Kepler Space Telescope yield a 49-day light curve with 98% coverage at a 1-minute cadence. A significant signature in the light curve comes from discrete cloud features. We compare results extracted from the light curve data with contemporaneous disk-resolved imaging of Neptune from the Keck 10-meter telescope at 1.65 microns and Hubble Space Telescope visible imaging acquired 9 months later. This direct comparison validates the feature latitudes assigned to the K2 light curve periods based on Neptune's zonal wind profile, and confirms observed cloud feature variability. Although Neptune's clouds vary in location and intensity on short and long time scales, a single large discrete storm seen in Keck imaging dominates the K2 and Hubble light curves; smaller or fainter clouds likely contribute to short-term brightness variability. The K2 Neptune light curve, in conjunction with our imaging data, provides context for the interpretation of current and future brown dwarf and extrasolar planet variability measurements. In particular we suggest that the balance between large, relatively stable, atmospheric features and smaller, more transient, clouds controls the character of substellar atmospheric variability. Atmospheres dominated by a few large spots may show inherently greater light curve stability than those which exhibit a greater number of smaller features.

1. Introduction

Brown dwarfs are substellar objects with masses below about 75 Jupiter masses, i.e., objects that cannot sustain hydrogen fusion (Chabrier et al. 2000).  Brown dwarfs share many aspects with giant planets; both classes are predominantly composed of hydrogen and helium with an admixture of other elements; both have cool (at least by stellar standards) atmospheres; both have atmospheres with molecules and condensates that strongly influence the transport of energy by radiation.  Review articles by Burrows et al. (2001), Burrows and Orton (2010), and Marley et al. (2013) compare and contrast the atmospheres of brown dwarfs and giant planets in more detail.

Many have searched for rotational and dynamical variability in brown dwarfs, dating back to shortly after their discovery (e.g., Tinney and Tolley 1999, Bailer-Jones and Mundt 2001, Gelino et al. 2002).  Recent studies reveal photometric variability of many brown dwarfs in the mid-infrared with the Spitzer Space Telescope (e.g., Metchev et al. 2015) and near-infrared spectral variability using the Hubble Space Telescope (e.g., Apai et al. 2013, Yang et al. 2015).  The most extensive ground-based survey was by Radigan et al. (2014), who found that L-type to T-type transition brown dwarfs are both more likely to be variable and show higher variability amplitudes than earlier and later spectral type objects.

While such variability has often been compared qualitatively to that seen in solar system giant planet atmospheres, there have not previously been truly comparable full-disk photometric studies for quantitative comparison.  Indeed, Radigan et al. (2014) reviewed the long history of variability searches in a variety of spectral bandpasses with a multitude of time baselines and sensitivities.  Despite the diversity in these searches, the unmistakable conclusion is that brown dwarfs are often variable.  Amplitudes ranged from a typical few percent to the current record of 26% variation in J band over about 8 hours by the T1.5 dwarf 2MASS J21392676+0220226 (Radigan et al. 2012).  This variability is typically attributed to inhomogeneous cloud cover resulting in a periodic brightness variation as the brown dwarf rotates.

A similar phenomenon of rotational modulation has been seen for giant planets in our own solar system, extending back over a hundred years to visual reports of planetary brightness modulations (e.g., Cassini 1665).  Ironically, even though the larger giants, Jupiter and Saturn, are resolved by even small telescopes, generating multi-rotation light curves is challenging due to their rotational periods versus ground-based visibility.  For these reasons it has been difficult to place the abundant brown dwarf variability data in the context of giant planet variability.  Gelino and Marley (2000) computed artificial visible and mid-infrared light curves for Jupiter by combining multiple full disk images, mapping them onto a sphere, and computing the expected rotational modulation in brightness.  Rotational modulation was maximized at IR wavelengths due to maximum contrast for large storms, like the Great Red Spot, suggesting that similar results would hold for brown dwarfs with patchy clouds (Karalidi et al. 2015).

To help fill this gap in light curve measurement of giant planets, our collaboration observed Neptune with the repurposed Kepler Space Telescope as part of the K2 extended mission (Howell et al. 2014). We chose Neptune because it is bright enough to extract a light curve with good photon statistics, but not so oversaturated that excess bleeding would substantially damage Kepler photometry (Stromgren b and y magnitudes of ~7.9 and ~7.75, respectively; Lockwood and Jerzykiewicz 2006).  In addition, it has exhibited clear rotational modulation in the past (e.g., Joyce et al. 1977, Lockwood et al. 1991).  Another key result from the Kepler prime mission was statistics of the size distribution of exoplanets, finding that hundreds were Neptune-sized (e.g., Batalha 2014). Thus, these observations provide ground truth for future photometry of exo-Neptunes (e.g., by space coronographs) and directly imaged exoplanets, in general, as well as brown dwarfs.

Kepler observations of Neptune were acquired from November 15, 2014 to January 18, 2015.  Neptune and its large moon Triton were visible with 98% coverage and a 1-minute observation cadence starting December 1, 2014, with Neptune subtending ~2.3" (a Kepler pixel covers ~4").  Due to Neptune's apparent motion, it crosses a 161x98 pixel subraster over the observational time frame (see the full transit video at:  https://www.youtube.com/watch?v=Tw-q3uM_5_0).  From this high cadence data set, we generate a 49-day high-precision light curve.  Kepler observes over visible wavelengths (e.g. Koch et al. 2010) from ~430 to 890 nm, and thus the light curve represents variations in Neptune's reflected solar flux, which necessarily combines variations both in Neptune's reflectivity and in the Sun itself.   Neptune, however, is a resolved object in ground and space-based facilities.  Thus, any inferred measurements from the light curves can be directly compared with observable discrete cloud features in the atmosphere, effectively providing ground truth for the Kepler light curve inferences.

In this paper, we described the results pertaining to Neptune's atmosphere, which dominates the Kepler light curve.  Separate papers will address the photometric signal from the Sun and the signal from Neptune's interior. We show correlation of the Kepler light curve output with contemporaneous Keck imaging and subsequent Hubble Space Telescope images, and compare with 20 years of Neptune cloud observations.  Short-term temporal evolution in the light curve is also addressed.  Finally, we discuss the implications for analyzing light curves of other potentially cloudy atmospheres.

2. About Neptune Light Curves

To first order, Neptune's rotational signature dominates the signature in the Kepler light curve, and stems from a few bright discrete features.  Such rotational modulation has been seen in light curves with far shorter baselines in the past (e.g., Lockwood et al. 1991).  Note that Neptune's internal rotation rate is in fact poorly defined, and was initially based on radio emissions detected by Voyager 2 that repeated every 16.11+/- 0.05 hours (Warwick et al. 1989).  Given only the brief flyby, it is still unclear if that represents a true core rotation rate; recent studies suggest that very stable polar cloud features may better constrain the rate to 15.96630 +/- 0.00003 hours (Karkoschka 2011).  For consistency with past publications, we adopt the usual value of 16.11 hours.

Assuming the 16.11-hour rotation rate, Voyager and subsequent ground-based observations showed that Neptune's apparent zonal winds vary with latitude (e.g., Sromovsky et al. 1995, 2001a, 2001b, Hammel and Lockwood 1997, Sanchez-Lavega et al. 2015).  Thus, Neptune light curves reveal differential rotation as features at various latitudes brighten and fade.  If a bright cloud feature moves with the local zonal wind, periodogram analyses can, to first order, be used to extract that cloud's latitude.  A subtlety is that sometimes Neptune's brightest features actually track large disturbances at other latitudes, e.g., the bright Companion Cloud to Neptune's Great Dark Spot was known to track the latitude of the dark feature, not the latitude of the bright companion itself (e.g., Smith et al. 1989, Sromovsky et al. 2011a).  Thus, caution must be exercised when extracting velocities from periodograms.

3. Kepler Light Curve Analysis

The raw Kepler data were processed by first subtracting the constant background star field using difference imaging and removing any remaining intrapixel variations.  Neptune saturates the CCD, but only to the level that adjacent pixels are illuminated by electrons that are transferred but not lost (e.g., Gilliland et al. 2001).  Thus, the signal can be summed into a disk-integrated value for each exposure. Periodic spacecraft motions, jitter, and reaction wheel desaturations are removed, using PSF photometry of the background stars.  To compute PSF models and determine photometric centroids we used methods developed for the MOST space telescope (Walker et al. 2003), specially designed for critically sampled, subrastered images of arbitrary size from space-based observatories (Rowe et al. 2006, 2008).  Lastly, we remove any residual bad points (spikes or drop outs).  These corrections result in photometry with a typical noise level of about 100 parts per million or better.

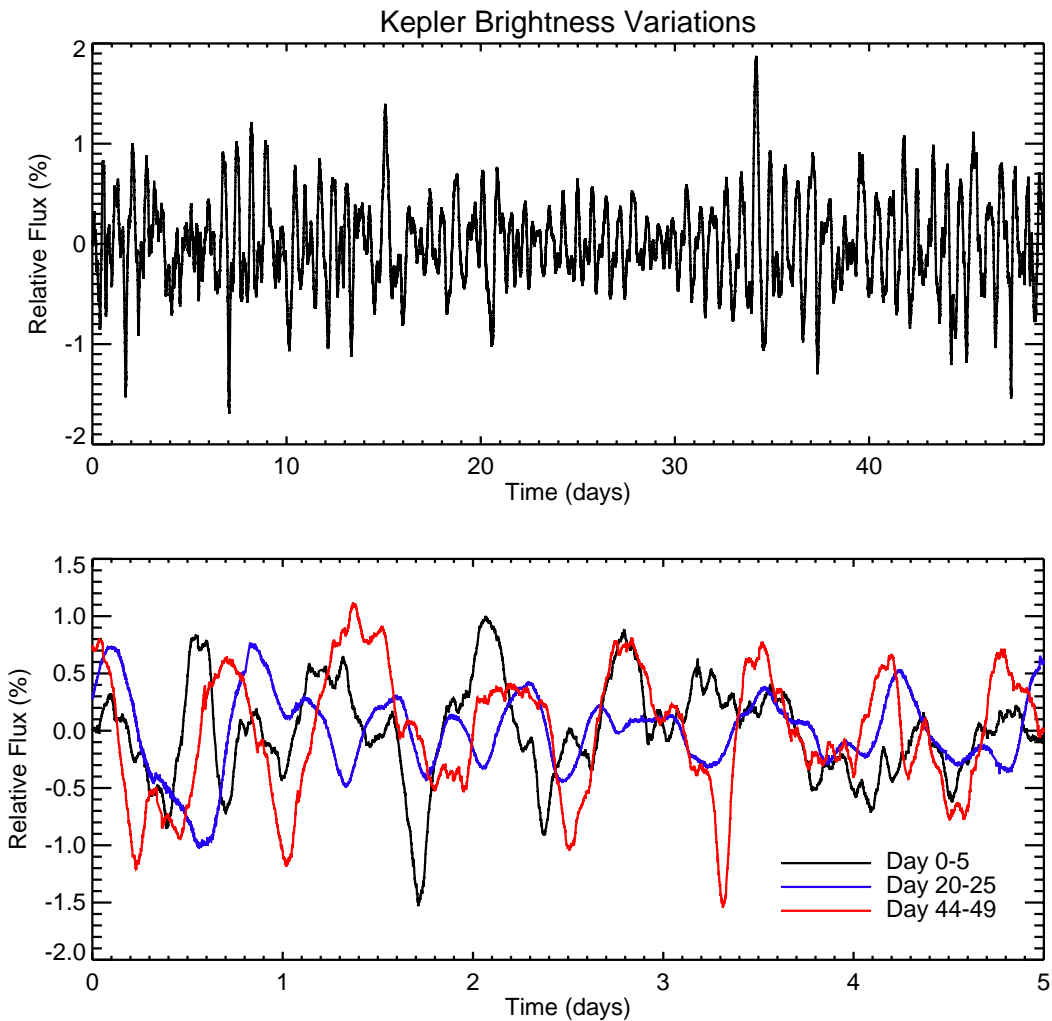

Figure 1. The Kepler light curve of Neptune. The top panel shows the full 49-day light curve, with normalized brightness variations over time elapsed since 1 Dec. 2014. The bottom panel shows several 5-day segments emphasizing the evolution of brightness variations with time.

The full data set includes 30-minute cadence data over a 70-day time period, but any remaining data discontinuities cannot be corrected at this cadence because real signals may be removed. However, the 49-day observations at 1-minute cadence allow for data discontinuities to be corrected. Figure 1 (top panel) shows the final extracted light curve as relative flux variations, after any remaining discontinuities and long-term trends have been removed. This curve shows a clear periodic signal, and a possible beat frequency, indicating more than one period is likely present. The curve is not perfectly smooth, with many small variations on top of the main signals. There is also some indication of time variability in the brightness and frequency of the variations (Fig. 1, bottom panel). This shows both the value, and the complexity, of a long-duration light curve covering ~73 rotations of the planet.

We analyzed the light curve in terms of frequencies by computing its power density spectrum with a fast Fourier transform, see Appendix A for details. We distinguish three groups of dominant peaks in between 15 and 17 uHz, 30 and 33 uHz, and 45 and 50 uHz, corresponding to Neptune's rotation and two of its harmonics. Figure 2 shows the periods corresponding to Neptune's rotation (15 to 17 uHz region). The three major peaks are undoubtedly signal (confidence level larger than 99.99%), while the neighboring peaks are marginally detected with confidence levels around 90% (in between 3 and 6-sigma).

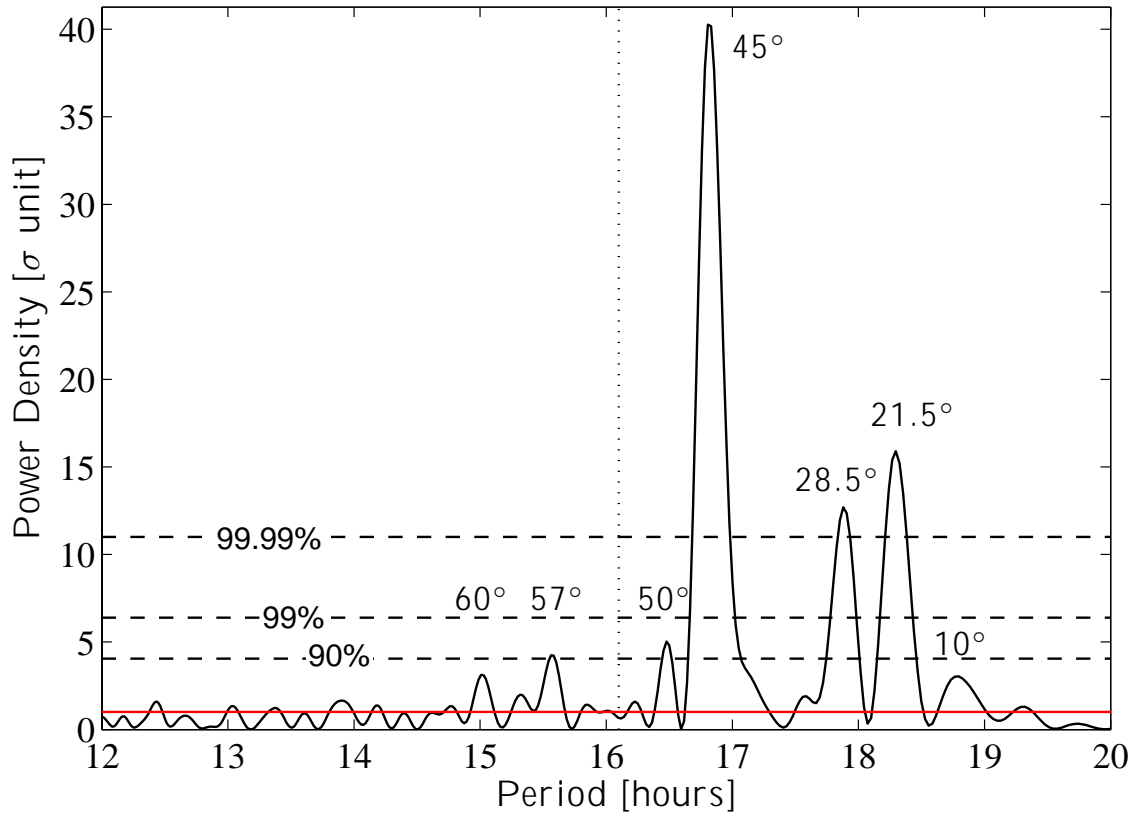

Figure 2. Oversampled and whitened power density spectrum as function of period. The red line corresponds to the noise level of the whitened spectrum and black dashed lines indicate statistical significance levels. Numbers above some peaks indicate the latitudes on Neptune corresponding to that rotation period based on the zonal velocity curve given by Sanchez-Lavega et al. (2015); the features could be in either hemisphere.

None of the peaks correspond to the periods of Neptune's major moons, nor their harmonics. Horizontal oscillations detected in prior Keck observations (Martin et al. 2012), potentially linked to tidal forcing by Triton, did not produce a corresponding 7.24-hour signal in our analysis of the photometric light curve. The peaks in the periodogram, if assumed to be created by discrete cloud features, can be used to infer the latitude of those features based on a symmetric zonal wind profile (Sanchez-Lavega et al. 2015). The most significant peaks roughly correspond to latitudes of 45°, 28.5° and 21.5° planetographic latitude, respectively. Since the wind profile is symmetric around the equator, these results cannot distinguish between northern or southern features, and we neglect any dispersion in the zonal velocities for the moment. We can break the hemispheric degeneracy with direct imaging observations of Neptune's cloud locations.

4. Neptune Cloud Activity During the Kepler Observations

We obtained disk-resolved imaging overlapping the K2 observations to provide ground-truth imaging for the photometry and to break the north-south degeneracy in the periodogram. Figure 3 shows rectilinear maps extracted from images obtained on 9 and 10 January 2015 with the Keck 2 10-meter telescope using the NIRC2 camera at H band (1.65 micron); this wavelength region is sensitive to relatively high clouds in the atmosphere, similar to visible red wavelengths, see Figure 4. Neptune typically shows less brightness variation at wavelengths shortward of 0.7 microns, therefore red and near-infrared wavelengths show most of the atmosphere's reflected light variability from distinct clouds. Past studies have shown that discrete clouds may be at altitudes as high as the 60-230 mbar pressure level, with the main methane haze/cloud layer near 1 bar pressure and clouds from other ices (e.g., $NH_3$, $H_2S$) possible at deeper levels (higher pressures) (e.g., Sromovsky et al. 2001a).

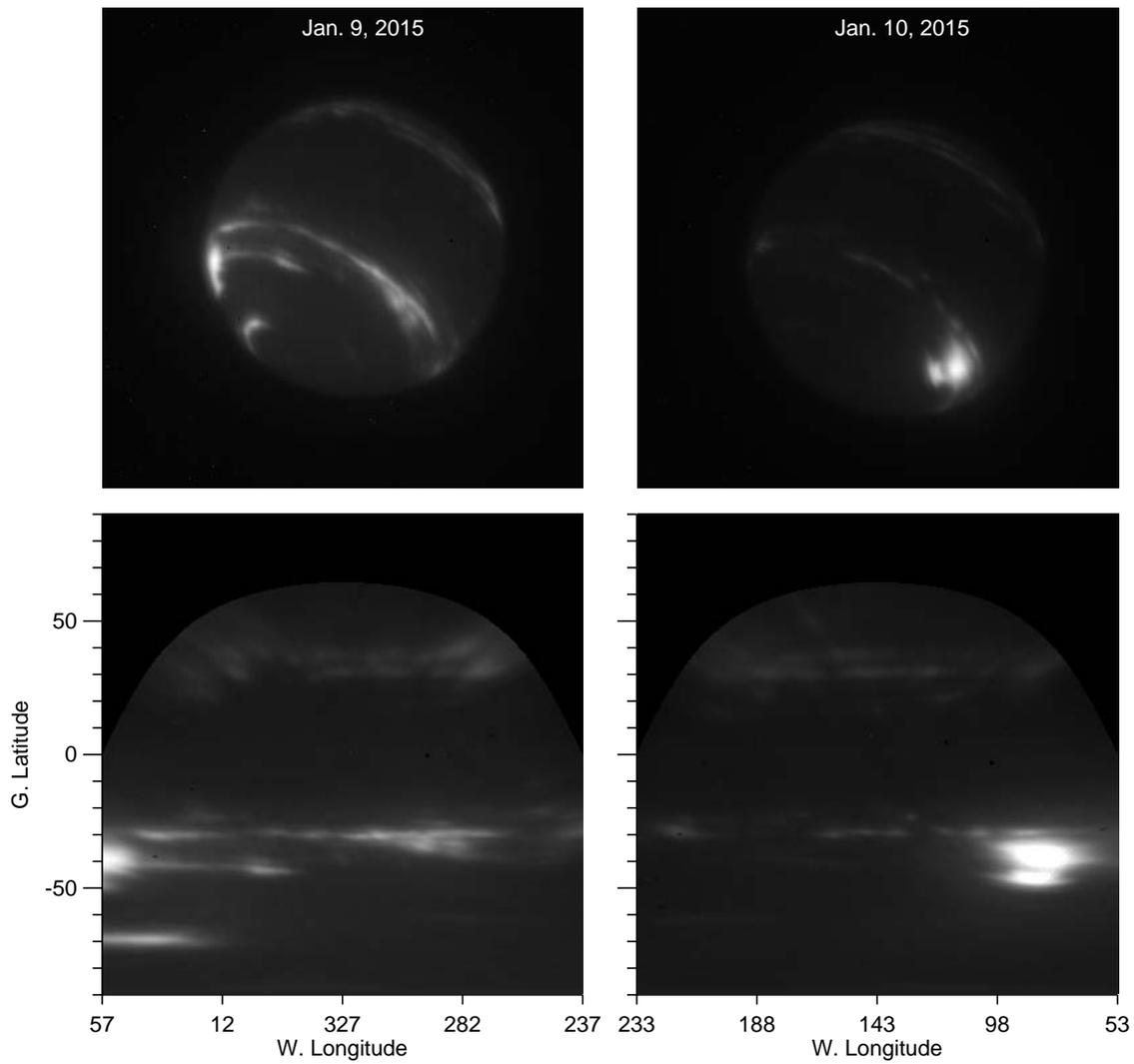

Figure 3. Keck H-band images of Neptune from 9-10 January 2015, covering most longitudes. The top panels are unmapped images, and the bottom panels show the latitude and longitude coverage mapped at 2 pixels per degree. These show typical Neptune structure: bright bands of Neptunian cloud activity from planetographic latitude 25° to 40° in the northern and southern mid-latitudes, with occasional brighter features.

A particularly bright discrete feature is seen at 80° W longitude in both images, although it is on the limb on the 9 January image. From this single image, one cannot tell whether this is a "complex" that extends over many latitudes but moves as one feature (e.g., the 1994 northern hemisphere complex; Hammel et al. 1995), or whether it is two separate features at 40° and 50° south that happen to align on this night. The very strong periodogram signature at a period corresponding to 45°, however, strongly suggests that this is indeed a "complex" that may correspond to a Great Dark Spot at 45° S, and that these bright features are companion clouds.

Another group of features that is bright and isolated enough to give a rotational signature is seen at 290° W on 9 January, extending from about 30°S to 45° S. Some of these features would also contribute to the periodogram signal at 45°. A steady smattering of features as function of longitude appears near latitude 28°S, which is consistent in the aggregate with the periodogram signature with that latitude.

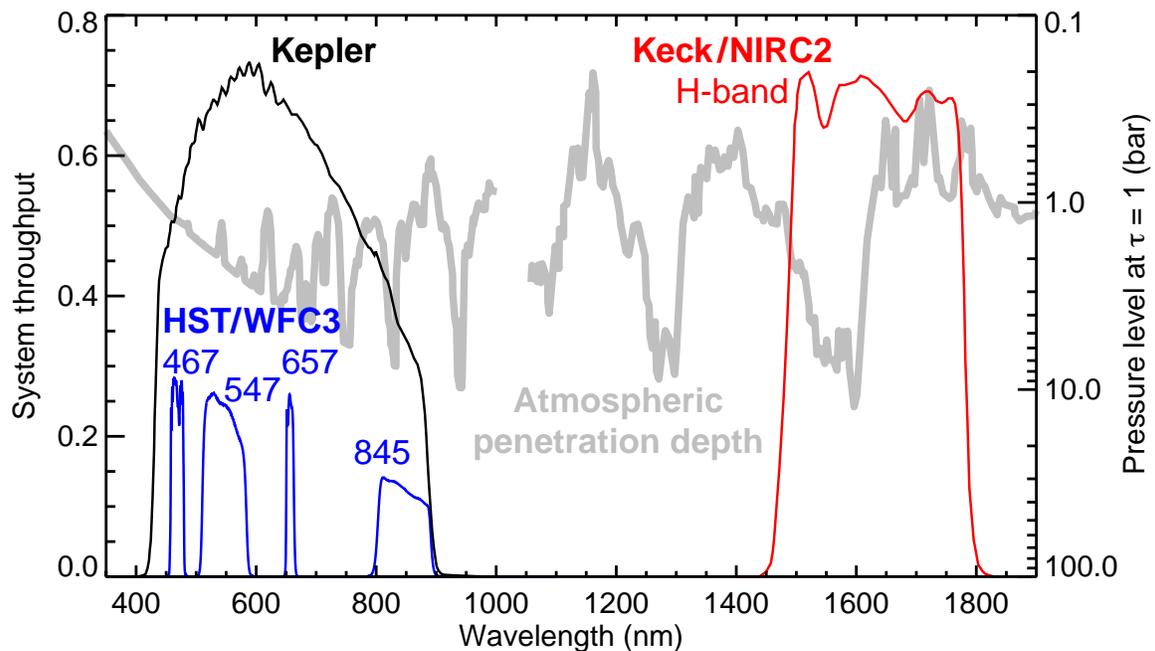

Figure 4. Spectral sensitivity and atmospheric transmission. Labeled curves show the total spectral sensitivity of Kepler and HST observations (Koch et al. 2010, Dressel 2015). The Keck infrared bandpass includes the NIRC2 H-band filter transmission and detector quantum efficiency, but neglects the telescope optical path outside NIRC2. The atmospheric penetration depth (right axis) is the pressure level where a two-way optical depth of unity is reached in a cloud-free model of Neptune's atmosphere, including opacity from Rayleigh scattering and gas absorption (from Sromovsky et al. 2001a).

The feature on 9 January at 50° west longitude (70° S) is likely the South Polar Feature (SPF) which has been imaged on numerous occasions (Smith et al. 1989, Rages et al. 2002, Karkoschka 2011). The rotation rate of this feature is quite stable at 15.97 hours (Karkoschka 2011), and does not match the zonal wind speed at this latitude, which has a period of 12.7 hours (Sanchez-Lavega et al. 2015). Its signature is not readily observed in the Kepler periodogram (Fig. 2), though its motion is consistent with the 15.97-hour period, as is discussed later.

It may not appear as discrete feature in the light curve due to its longitudinal extent and Neptune's inclination (the sub-Earth latitude is 27° South on these dates), making the SPF visible for much of a rotational period.

Regarding the remaining features in the periodogram and their presumed latitude, there is no obvious corresponding cloud feature near 20° N or S. The Keck data were acquired near the end of the Kepler 49-day time frame, so it is possible that features may have evolved in brightness or migrated in latitude over the Kepler time frame. Additionally, the mean wind profile may not be an accurate representation of the velocity of the visible features, as is noted for the SPF. Lastly, these near-infrared images do not represent the visible-wavelength appearance of the planet (which senses a lower altitude) from which we derive the light curve; more cloud features are likely visible in the high contrast Keck data, based on the atmospheric transmission curve in Fig. 4.

5. Other Neptune Observations

Hubble data were also acquired in September 2015 as part of the "Hubble 2020: Outer Planet Atmospheres Legacy" (OPAL) program (Simon et al. 2015). The OPAL program generates two global Neptune maps each year using the Wide Field Camera 3 (WFC3). A main goal of OPAL is to provide Neptune data for long duration time-domain studies of cloud activity and wind field variability, making it a perfect companion to this work. Although the data were acquired well after the Kepler observations, they enable an independent high-spatial-resolution look at the clouds at visible and near-IR wavelengths to show how much they vary over 9 months.

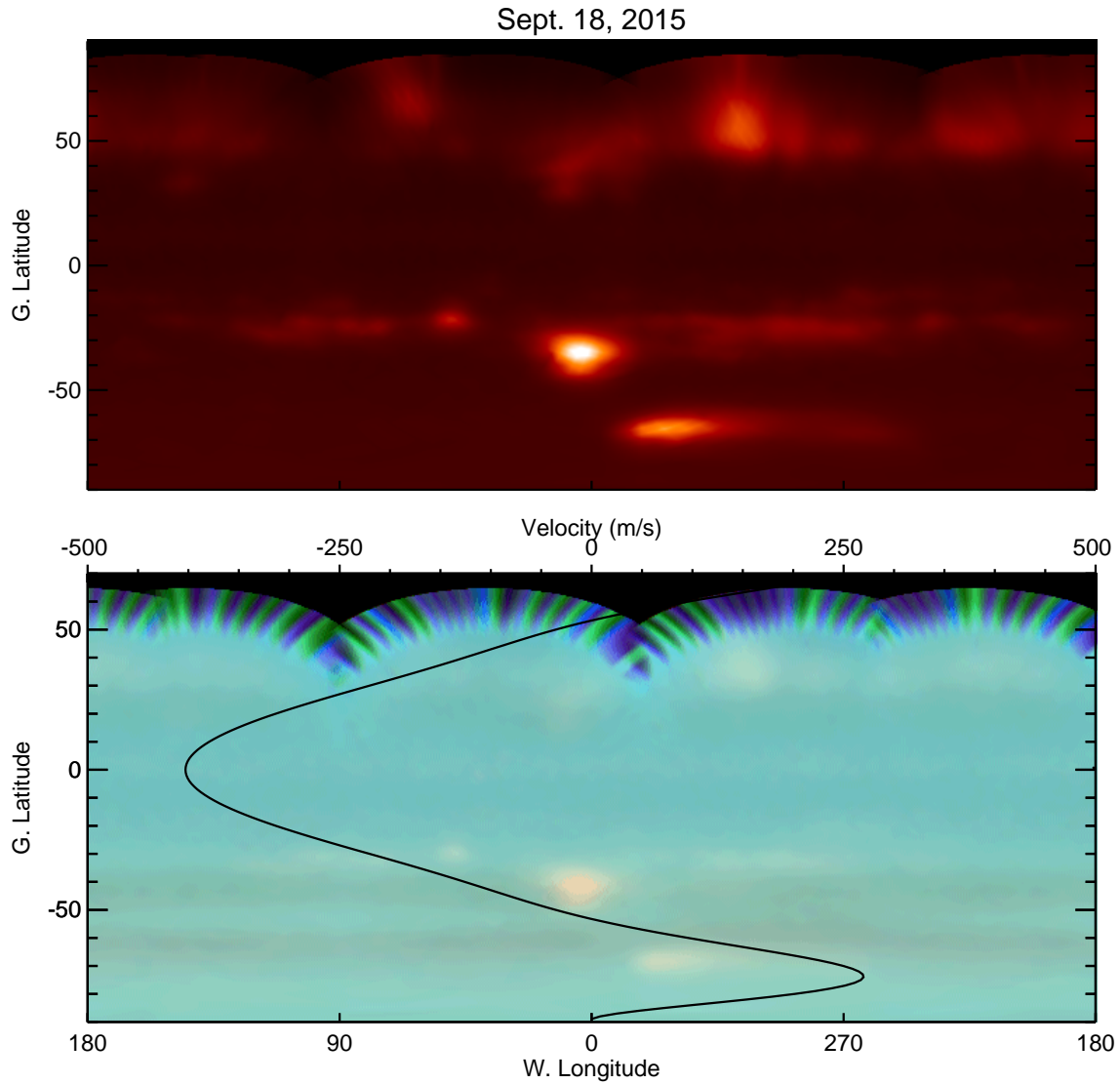

Figure 5. Hubble map of Neptune acquired 18 September 2015. The top panel shows a global map constructed from 845-nm images. The bottom is a visible-wavelength color-composite map (with the blue, green, and red channels mapped to 467, 547, and 657 nm, respectively). We overplot the smoothed zonal wind profile (Sanchez-Lavega et al. 2015), showing winds up to 400 m/s (top axis).

Figure 5 shows Hubble observations of a complete rotation of Neptune, created from 4 orbits. Very similar cloud morphology is seen in the Keck H-band map (Fig. 3) and the Hubble 845-nm map (Fig. 5, top), including the large storms system near latitude 45° S and the bright SPF at latitude 70° S. However, fewer features are observed near 25° N, implying some variability since January 2015. The color comparison (bottom panel in Fig. 5) shows that many of the cloud features are muted at shorter visible wavelengths, and darker bands also appear from 40° to 50° S and from 60° to 70° S.  Thus, a panchromatic visible light curve would be dominated by the variable clouds at the longer wavelengths (i.e., by the features that appear white in the composite).

Observations of a second rotation of the planet were not completed due to a spacecraft tracking anomaly; only part of the second map was obtained leaving a longitude gap from 235 to 308° W.  However, many of the cloud features were captured, allowing for feature motion measurements; these generally match the wind profile in Fig. 5, with the exception of the SPF.  Small variations are expected, as larger cloud features can also have internal rotation and drift rates that do not represent the mean zonal wind.  This is particularly true of the SPF, which drifts at a much slower rate than the zonal wind at that latitude.  Previous cloud motion measurements indicate velocity dispersions of 200 m/s or higher, indicating much variability in feature motions; it is unclear if this also applies to the true zonal wind field as feature motions may not be identical to the zonally averaged wind (Martin et al. 2012, Fitzpatrick et al. 2014).

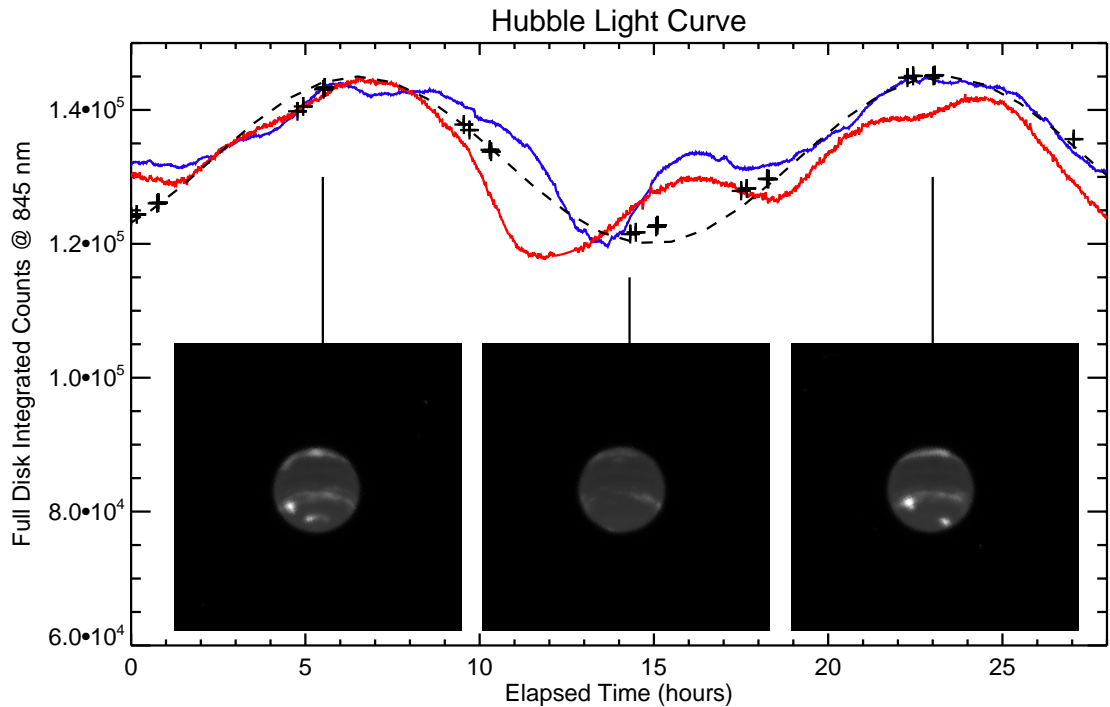

Figure 6.  Light curve of Neptune from Hubble full-disk brightness at 845 nm (plus signs). A sinusoidal variation, with a 16.8-hour period and arbitrary amplitude, is shown by the dashed line.  For comparison, normalized Kepler light curves beginning at Day 6 and Day 25 are shown in blue and red, respectively.

In addition, the 845-nm filter was sampled repeatedly within the orbits, giving additional coverage of features and full disk measurements.  Some small changes in cloud morphology were observed, but these are unlikely to affect a disk-integrated light curve. From the 25 exposures obtained in the 845-nm filter, we extracted the full-disk brightness to generate the light curve shown in Figure 6 (see the Supplemental Online Material for a full animation of the images and light curve). Although a periodic signal with a minimum to maximum amplitude of ~16% can be seen in this light curve, a Lomb-Scargle periodogram cannot pull out a unique period because of the sparse temporal coverage. The dashed line indicates the 16.8-hour period expected if the 45° S feature dominated this light curve, and red and blue curves show normalized Kepler light curves from Days 6 and 25, respectively. The Hubble light curve represents the maximum variation we would expect to see, as cloud contrast is maximized. Full disk counts in the 467-nm filter, from the darkest to brightest views, give a 0.2% variation in total integrated counts, at the limit of the WFC3 photometric accuracy (Kalirai et al. 2009, 2010).

The smaller cloud signal observed at shorter wavelengths is due to the atmospheric levels sensed by these filter bandpasses, as shown in Fig. 4. The shortest wavelengths are dominated by Rayleigh scattering, which gives an overall bright atmospheric background, reducing contrast for discrete cloud features. At longer wavelengths, Rayleigh scattering is reduced and particle scattering above the 1-bar pressure level can be more easily detected. At methane and other gas absorption bands, photons are absorbed before reaching deeper cloud levels, and higher clouds show high contrast from the rest of the atmosphere, for example at 890 nm. Thus, at shorter wavelengths, or with a panchromatic visible bandpass, the light signal from discrete clouds is much more muted than at red and infrared continuum or absorption band wavelengths.

6. Discussion

The data acquired in 2015 from Keck and Hubble show that the planet varies on a timescale of hours to months. The largest feature at 45° S has been quite stable, however, as have the location of some of the bands of cloud activity. On the other hand, the planet can show dramatic variability in clouds. Figure 7 shows a similar map from Hubble data acquired in 2011 at 845 nm. Here there are no complete bands of clouds, but many more discrete clouds. During the Voyager 2 flyby in 1989 there were few bright clouds, and Neptune's dominant cloud features were the Great Dark Spot near ~15° S, the SPF near 70° S, another dark spot near 55° S and a bright cloud near 45° S (Smith et al. 1989). However, Neptune's more usual appearance includes bands of activity with discrete storms. Table 1 provides an selected sampling of cloud activity on Neptune over the past 20 years to show that some latitudes have fairly constant cloud activity, but many more evolve with time. In several of these cases, cloud evolution was seen over just a few days or even hours (Sromovsky 2001b, Fitzpatrick 2014).

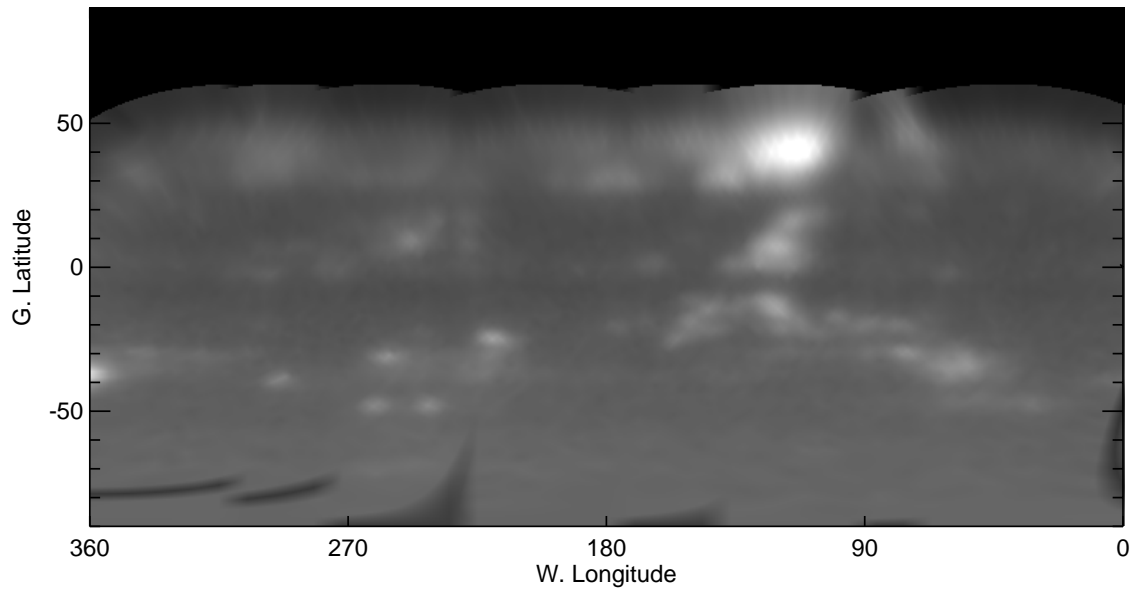

Figure 7: Neptune global map from Hubble WFC3/UVIS acquired 25-26 June 2011 at 845 nm. High northern latitudes were not visible, and a bad column resulted in artifacts at high southern latitudes; no SPF is visible.

Table 1. A limited 20-year summary of cloud detections

| Date | Facility | North | South | Reference |
|---|---|---|---|---|
| 1994 | Hubble | Discrete dark feature at 30°, bright features at 27° N to equator | Bright features at 30° and 45° S | Hammel et al. 1995 |
| 1996 | Hubble/ NASA's IRTF | Discrete features near 20° to 40° N | Bands near 20° to 40° w/ features, feature near 60° S | Sromovsky et al. 2001a |
| 1998 | Hubble | Features between 20° and 50° N | Band near 45°, features at 15° to 40° S | Sromovsky et al. 2001b |
| 2001 | Keck | Band at 28° N, sporadic bright clouds at 36° N | Bands at 23°, 31°, 36°, 45° and 49° S | Martin et al. 2012 |
| 2001 | Hubble |  | Bright feature at 70° S | Rages et al. 2002 |
| 2003 | Keck, VLA | Bands between 25° and 40° N | Bands between 30° and 50° S, discrete features near 60-70° S | de Pater et al. 2014 |
| 2009 | Keck | Bands at mid | Large feature at 65° | Fitzpatrick et |

| | | latitudes, features at 40° N | S | al. 2014 |
|---|---|---|---|---|
| 2011 | Hubble | Broken bands, many features | Broken bands, features from 10 to 50° S | this paper |
| Jan. 2015 | Keck | Bands from 25° to 40° N | 30° S, features at 40° to 50° S | this paper |
| Sep. 2015 | Hubble | Broken band 30° to 40° N, features at 20° N | Bands from 25° to 40° S, features at 45° and 70° S | this paper |

In addition to changing cloud activity, Neptune's longer-lived features can oscillate in latitude and longitude. Voyager 2 images showed that features near 21°, 42° and 54° S latitude could oscillate by 2° to 4° latitude and 8° of longitude (Sromovsky 1991). With the long Kepler coverage, it is possible that different periods, corresponding to different latitudes, could be found if binned over smaller time intervals rather than searched over the entire 49-day duration. Figure 8 shows Lomb-Scargle periodogram analyses run over 3.5-day intervals (5.25 Neptune rotations).

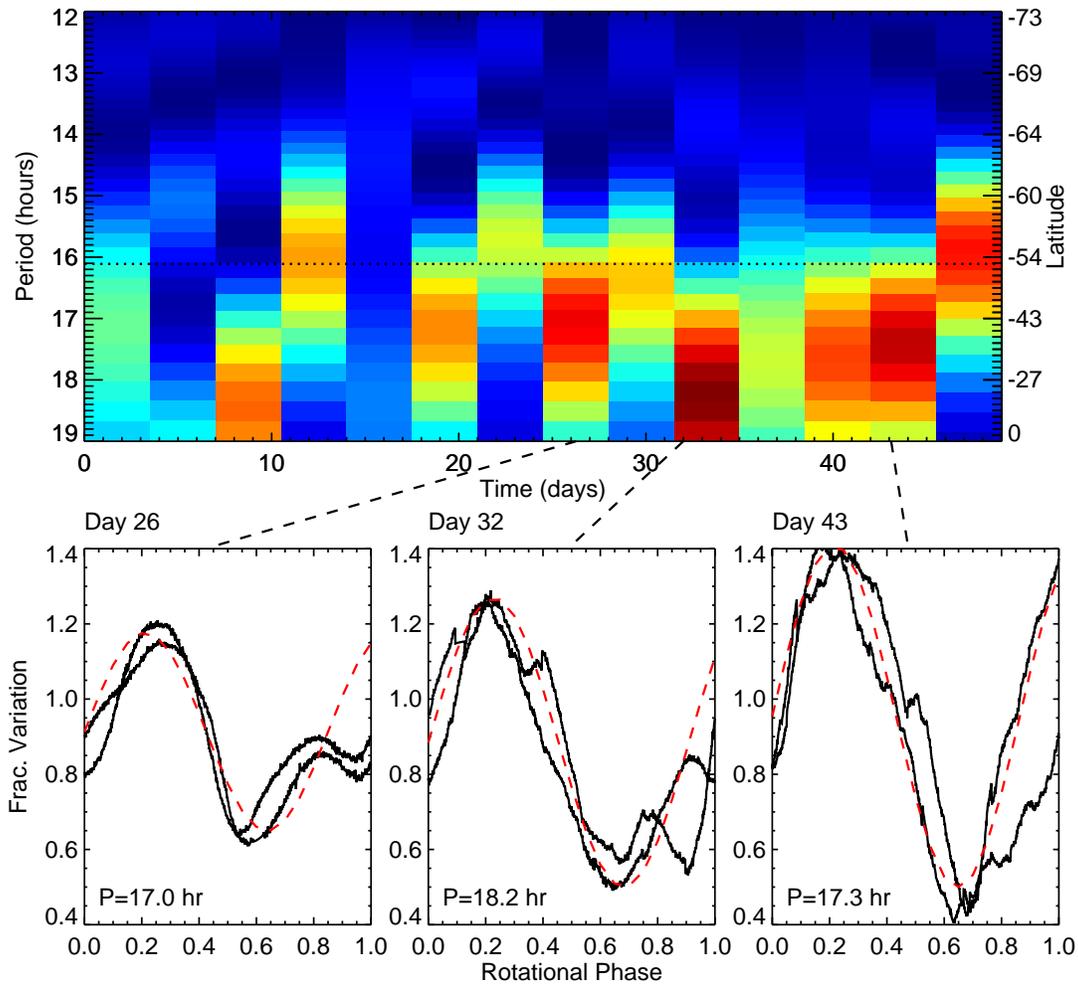

Figure 8. Short-interval periodogram analysis. The top panel shows the Lomb-Scargle periodogram in 3.5-day segments; red indicates higher spectral power. The remaining panels show the Kepler brightness variations (black curves) from three of the segments, rotationally phased to the corresponding period of maximum spectral power from the periodogram, and plotted over two rotations within that interval; the most significant period is shown as a dashed red line for each date.

Spectral power peaks are seen in every segment, but none show multiple peaks, and the variations are too large to represent a single cloud feature's motion. Rather, different features may dominate on different days, as they brighten or spread and then dissipate. For example, the signature of the 70° S feature may be dominating the signal at Days 45-49, even though it is not seen in the full periodogram in Fig. 2. This is not unusual, as observations from Keck and Hubble over 2000 to 2001 showed that the SPF feature can come and go, evolving on timescales as short as 30 hours and visible in about 20% of observations (Rages et al. 2002). Additionally, noise may be preventing clean retrievals of multiple features over so few rotations of the planet.

The observed Neptune variability has implications for brown dwarf light curve analyses. While some brown dwarfs show remarkably consistent light curves (e.g., Gizis et al. 2015 and examples cited therein), the light curves of other brown dwarfs evolve with time. In their study with the Spitzer Space Telescope of photometric variability of L3-T8 dwarfs, Metchev et al. (2015) found that about half were variable in IRAC bands 1 and 2 and of these about 1/3 showed rapid light curve evolution (over timescales of hours).

The largest amplitude variability among brown dwarfs occurs at the L to T type transition, where the thick cloudy atmospheres of the late L dwarfs transition to the relatively cloud-free spectra of the mid to late T dwarfs. For example, the J band thermal emission of the T2.5 brown dwarf SIMP J013656.57+093347.3 shows peak-to-valley variations as large as 5% with a period of a few hours (Artigau et al. 2009). The dwarf's light curve clearly evolves with time, exhibiting clear morphological differences in a few dozen rotations. Artigau et al. (2009) attribute the variations to evolution of patches of clear and cloudy regions in the atmosphere. Likewise Radigan et al. (2012) found large (26%) variations in the JHK thermal flux from the T1.5 dwarf 2MASS J21392676+ 0220226 with a period of about 8 hours. The light curve shape of this object also evolves over a few rotation periods, and Radigan et al. also attribute this to evolving photospheric clouds.

In perhaps the best known example of T dwarf variability, Gillon et al. (2013) monitored the L7.5/T0.5 binary WISE J104915.57-531906.1, commonly known as Luhman 16AB. They found 11% variability in the atmosphere of the cooler (T0.5) component that notably evolved over 12 nights of observations. Crossfield et al. (2014) later used Doppler imaging techniques to resolve individual bright and dark spots over the disk of the T dwarf, supporting the interpretation that photospheric clouds were responsible both for the periodic modulation of the light curve as well as its evolution in time.

It is interesting to consider the light curve evolution of Neptune in this context. First, it is worth repeating that the Neptune variability detected by K2 arises not from variations in the thermal flux but rather from variations in the reflectivity of the global cloud deck (although temperature contrasts within the atmosphere may well play a role in the evolution of the cloud features). The main component of the Neptune light curve (Fig. 1) is the dramatic bright spot. This feature is long lived and is responsible for the principal component of the variation over a single rotation (Figs. 2 and 7). However, multiple smaller features both produce irregularities in the light curve and seem to evolve over more rapid timescales, at time as quickly as within a rotation or two (Fig. 8). Without the largest feature, the light curve would be far more irregular, and without the varying smaller spots the rapidity of the evolution would be much less.

Stellar spot modeling (e.g., Mosser et al. 2009, Karalidi et al. 2015) can extract latitude information depending on the stellar inclination combined with assumptions about spot size and albedo.  At 90° inclination, no transits/modulations are seen, and at zero inclination, all spots transit in half a rotation period; other inclinations allow reasonably constrained retrieval of spot latitude to +/- 10° to +/- 20° (Mosser et al. 2009).  As Neptune has a tilt of 26° during these observations, this type of spot-latitude modeling would provide an interesting comparison to our work.

It should also be noted, however, that Neptune has large, latitude-dependent zonal wind velocities of several hundred meters per second, and some clouds move at the corresponding zonal velocity, while others do not.   Without prior knowledge of Neptune's zonal wind field, we could not assign latitudes to any particular period in the light curve, and no features appeared at the presumed internal 16.11-hour rotation period.   For comparison, Jupiter has lower maximum zonal velocities (~150 m/s), lower obliquity (3°), and its storms typically drift at lower velocity than the corresponding zonal winds (e.g., Beebe et al. 1989, Simon and Beebe 1996).  Here, modeling a short duration light curve does extract Jupiter's rotation rate and Great Red Spot latitude, though other spots are not obvious due to small size, low contrast, and degeneracy in the latitude retrievals (Karalidi et al 2015).  In principle, longer cadences could provide some zonal wind information, at least for latitudes with high contrast, distinct, cloud features, though they will be biased by the storm's own motions.  This highlights the importance of simultaneous resolved imaging when possible.

Perhaps the diversity seen among brown dwarf light curves—with some exhibiting relatively stable sinusoidal variations while others show either no regularity or rapidly evolve—is likewise a consequence of the balance of large, high-contrast features with smaller, more dynamic features. A logical next step would be to compare the observed Neptune variations to the predictions of a global climate model that could investigate the atmospheric dynamics both of irradiated giant planets and brown dwarfs, as well as to study long-duration light curves from the other solar system giant planets.  A statistical study of the types of weather patterns and their resulting variability would inform discussions such as these.

Acknowledgements
This paper includes data collected by the Kepler mission, available through the MAST archive: http://archive.stsci.edu/k2.  Funding for the Kepler mission is provided by the NASA Science Mission Directorate. We acknowledge T. Barclay for assistance with the K2 data reductions.  This work was based, in part, on observations made with the NASA/ESA Hubble Space Telescope under programs GO12675 and GO13937.  Support for program GO13937 was provided by NASA through a grant from the Space Telescope Science Institute, which is operated by the Association of Universities for Research in Astronomy, Inc., under NASA contract NAS5-26555.  Some of the data presented herein were obtained at the W.M. Keck Observatory, which is operated as a scientific partnership among the California Institute of Technology, the University of California and the National Aeronautics and Space Administration. The Observatory was made possible by the generous financial support of the W.M. Keck Foundation.  We thank D. Piskorz, H. Ngo, and H. Knutson for acquiring the Keck images of Neptune.  The authors wish to recognize and acknowledge the very significant cultural role and reverence that the summit of Mauna Kea has always had within the indigenous Hawaiian community.  We are most fortunate to have the opportunity to conduct observations from this mountain.

Appendix A: Computing the FFT Periodogram and Significance

To compute the FFT shown in Fig. 2, short interruptions in the dataset were filled by interpolating with the neighboring bins.  The resulting power spectrum looks very similar to those of the Sun and solar-like stars (e.g., Chaplin et al. 2010), with groups of peaks corresponding to Neptune's rotational features (15 to 17 uHz) and its harmonics (30 to 33 uHz, and 45 to 50 uHz).  We fitted the noise level with a maximum likelihood estimator to then whiten the power spectrum, Figure 9.  The global noise level can be easily modeled by a sum of two semi-Lorentzians plus the white noise level (e.g., Appourchaux et al. 2006). To better determine the peak frequencies (i.e., rotation periods), we oversampled the power density spectrum by a factor ten.  We then applied a statistical null hypothesis testing (H0) with the method proposed by Gabriel et al. (2002) for oversampled spectra to determine significance levels, as shown in Fig. 2.  A Lomb-Scargle analysis performed without data gap filling, spectral oversampling, or noise whitening, gives identical peaks, but the significance estimation is less robust.

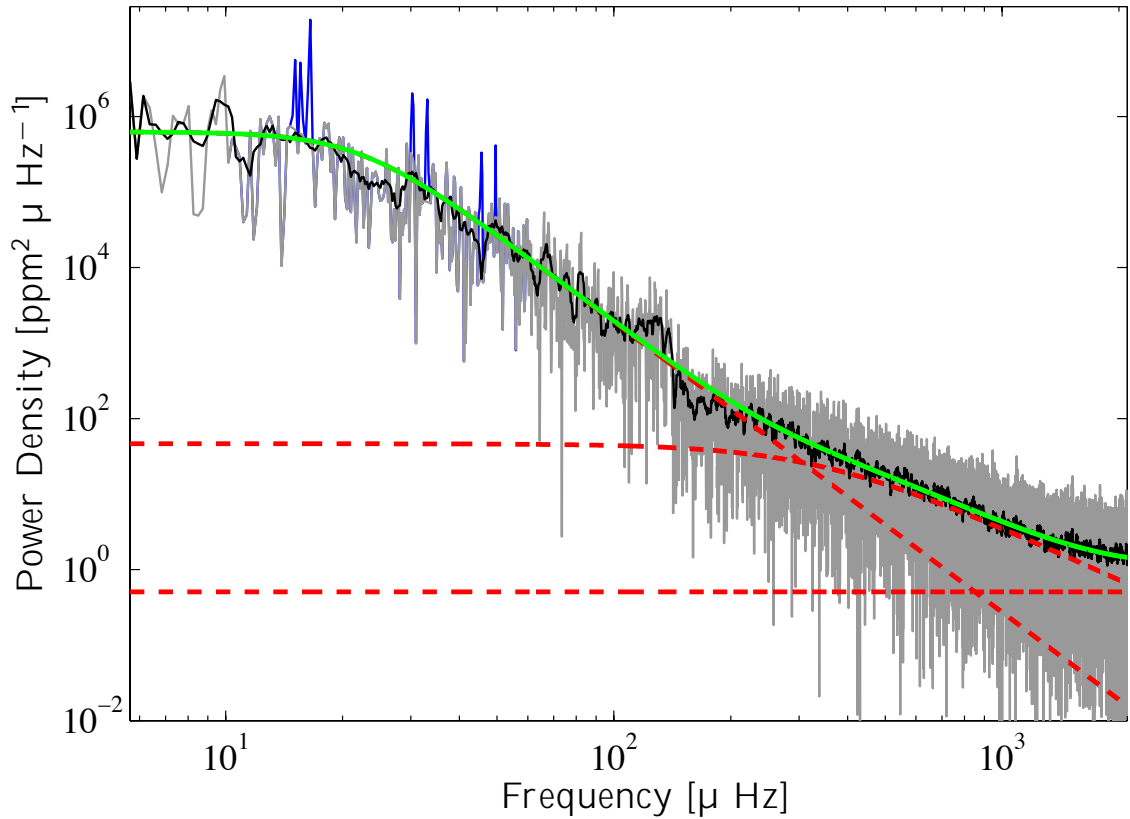

Figure 9. Fourier transform of the Kepler light curve. Grey line is the power density spectrum of the Kepler light curve in parts per million (ppm) per uHz, as function of frequency (uHz). The blue peaks match Neptune's rotation frequency and two harmonics. The black line is the power density smoothed over 100 bins to guide the eye to the mean noise level. The plain green line indicates the noise model, which is the sum of two semi-Lorentzians and a white noise offset (dashed red lines).

Supplemental Movie: Time series of Hubble 845-nm images and their corresponding light curve. These images were acquired in September 2015, and although separated from the Kepler data by more than 9 months, the same large cloud system is present. The dominant 16.8-hr period found in the Kepler data is overplotted for comparison, dashed line.